\documentclass{aa}

\usepackage{natbib}
\bibpunct{(}{)}{;}{a}{}{,} 

\usepackage{graphicx}
\newcommand{\gsim}{\,\raisebox{0.2em}{$>$}\!\!\!\!\!
\raisebox{-0.25em}{$\sim$}\,}
\newcommand{\lsim}{\,\raisebox{0.2em}{$<$}\!\!\!\!\!
\raisebox{-0.25em}{$\sim$}\,}
\newcommand{\gr}{$\gamma$-ray \,}

\begin{document}
\title{Gamma-ray emission from Cassiopeia A produced by accelerated cosmic rays}
\titlerunning{Gamma-rays from Cas\,A produced by accelerated CRs}

\subtitle{}

\author{E. G. Berezhko
        \inst{1}
        \and
          G. P\"uhlhofer
          \inst{2}
         \and
         H. J. V\"olk
         \inst{2}}

\offprints{H. J. V\"olk}

\institute{Institute of Cosmophysical Research and Aeronomy,
                     31 Lenin Ave., 677891 Yakutsk, Russia\\
              \email{berezhko@ikfia.ysn.ru}
         \and
             Max Planck Institut f\"ur Kernphysik,
                Postfach 103980, D-69029 Heidelberg, Germany\\
             \email{Gerd.Puehlhofer@mpi-hd.mpg.de}
             \email{Heinrich.Voelk@mpi-hd.mpg.de}
             }

\date{Received month day, year; accepted month day, year}

\abstract{
The nonlinear kinetic model of cosmic ray (CR) acceleration in supernova
remnants (SNRs) is used to describe the relevant properties of 
Cassiopeia~A (Cas\,A). In order to reproduce the SNR's observed size, expansion rate
and thermal X-ray emission we employ a piecewise homogeneous model for the
progenitor's circumstellar medium developed by 
\citet{Borkowski_ApJ_1996_466}.
It consists of a tenuous inner wind bubble, a dense shell of swept-up red
supergiant wind material, and a subsequent red supergiant wind region. A
quite large SNR interior magnetic field $B_\mathrm{d}\approx 1\,\mathrm{mG}$ is required
to give a good fit for the radio and X-ray synchrotron emission. The steep
radio spectrum is consistent with efficient proton acceleration which
produces a significant shock modification and leads to a steep electron
spectrum at energies $\epsilon_{\mathrm{e}}<1\,\mathrm{GeV}$. The calculated integral
$\gamma$-ray flux from Cas\,A, $F_{\gamma}\propto \epsilon_{\gamma}^{-1}$,
is dominated by $\pi^0$-decay $\gamma$-rays due to relativistic protons.
It extends up to roughly $30\,\mathrm{TeV}$ if CR diffusion is as strong as the Bohm
limit. At TeV energies it satisfactorily agrees with the value 
$5.8\times 10^{-13}\mathrm{cm}^{-2}\mathrm{s}^{-1}$ detected by the HEGRA collaboration.

\keywords{supernovae: individual: Cassiopeia\,A -- cosmic rays -- gamma rays: theory
-- acceleration of particles -- shock waves -- radiation mechanisms: non-thermal}
}

\maketitle

%

\section{Introduction} 
 
Cassiopeia~A (Cas\,A) is a prominent shell type supernova remnant (SNR),
and a bright source of synchrotron radiation observed at radio frequencies
\citep[e.g.][]{Bell_Nature_1975_257,Tuffs_MNRAS_1986_219,Braun_AA_1987_171,Andersen_ApJ_1991_373,Kassim_ApJ_1995_455}
and most probably also in the X-ray band 
\citep{Allen_ApJ_1997_487,Favata_AA_1997_324}.
Since SNRs are widely suspected to be
the main sources of the Galactic cosmic rays (CRs) up to energies of at least
$10^{15}\,\mathrm{eV}$, this youngest Galactic SNR offers
an important possibility to test this CR origin hypothesis. 
With the synchrotron interpretation,
the measured emission of hard X-rays is direct evidence for
the existence of a large number of relativistic electrons with energies up to
about $10\,\mathrm{TeV}$, presumably accelerated at the outer SNR shock.
Experimental information about nuclear
CR production can only be obtained by high energy \gr measurements. If
protons are accelerated in Cas\,A to at least the same energy and as
efficiently as electrons then the $\pi^0$-decay \gr spectrum, created in
their hadronic collisions with the background nuclei, should extend to
energies above $1\,\mathrm{TeV}$ with a hard power-law. The detection of a signal in
TeV $\gamma$-rays has been indeed recently reported by the HEGRA
collaboration \citep{Aharonian_AA_2001_370}. Therefore it is of the 
exceptional importance to find out whether this \gr emission is consistent with a
hadronic origin.

In this paper a thorough theoretical analysis will be given, based
on earlier work \citep{Berezhko_ProcICRC_2001_2}. It follows similar
analyses for SN\,1006 \citep{Berezhko_AA_2002_395} and for Tycho's SNR
\citep{Voelk_AA_2002_396}. We note that in the meantime also new
$\gamma$-ray observations of SNR\,RX\,J1713.7-3946 by the
CANGAROO experiment have been discussed \citep{Enomoto_Nature_2002_416} and
criticized \citep{Reimer_AA_2002_390, Butt_Nature_2002_418} on a
phenomenological basis. Unfortunately, the latter object is up to
now only very poorly understood, for example regarding the type of
explosion (quasi-explosive nuclear burning of an accreting White
Dwarf vs. core collapse of a massive progenitor star) and the
distance (estimates range from 1\,kpc to 6\,kpc). The object also
lies in an environment of complex morphology. In addition the
radio observations are so scarce that they do not allow to
determine a spectrum. As a result only phenomenological single 
box-type estimates for the particle spectra which are responsible for the
synchrotron/Inverse Compton radiation and
the hadronic $\pi^0$-decay $\gamma$-ray emission have been made up to now,
preventing any firm conclusion regarding the nature of the
$\gamma$-ray emission. Even a detailed theoretical model like ours
could, with the currently available data,
only lead to a number of constraints. In contrast, Cas\,A is
perhaps the best-studied object of its kind. The following
analysis makes detailed use of this impressive multi-wavelength
knowledge.

The chemical compositions of characteristic parts of the supernova (SN) ejecta 
\citep[the fast moving knots, e.g.][and references therein]{Reed_ApJ_1995_440}
and of the circumstellar material 
\citep[the quasi-stationary flocculi,][]{Peimbert_ApJ_1971_170,Peimbert_ApJ_1971_167,Kirshner_ApJ_1977_218,Chevalier_ApJ_1978_219}
suggest that the progenitor star started as a massive
object. It should have evolved from a hot main sequence star with a 
tenuous fast wind
to a red supergiant phase with a slow but dense wind, and finally into a
hot Wolf-Rayet star with a fast wind again, sweeping up part of the
preceeding red supergiant material into a dense shell before exploding as
a core collapse SN 
\citep{Chevalier_ApJ_1989_344,Garcia-Segura_AA_1996_316}.
In this spirit 
\citet{Borkowski_ApJ_1996_466} 
modeled the thermal X-ray emission as
well as the size and the expansion rate of the present SNR from their
result that the outer blast wave had already passed the swept-up red
supergiant wind shell and was presently propagating through the
unperturbed slow red supergiant wind. According to the X-ray measurements of 
\citet{Favata_AA_1997_324}
the SNR shock has swet up about 5$M_{\odot}$ of this
circumstellar material. 
Therefore we can expect a detectable level of $\pi^0$-decay \gr flux.

Empirical arguments for such a structure of the circumstellar material
around the
presupernova star come for example from the radio observations of Cas\,A.
Despite its youth of about 320 years, characterized also by the
relatively small amount of matter swept up by the SN shock,
the observed radio flux undergoes considerable secular decline 
\citep[e.g.][]{Rees_MNRAS_1990_243},
which amounts to about 0.6\% per year. For a uniform 
circumstellar material one would expect 
the radio flux to increase until an intermediate Sedov
stage of evolution, corresponding to an age of about $10^3\,\mathrm{yr}$. Only in
the case when the SN shock propagates through a wind of the progenitor
star with progressively decreasing density $\rho\propto r^{-2}$, the total
radio emission is expected to decrease with time 
\citep[e.g.][]{Berezhko_Jetph_1999_89}.

This suggests two rather complementary approaches regarding the production
of nonthermal particles and their radiation. One can either emphasize the
strong radio emission from small inhomogeneities whose free energy
(kinetic and magnetic) is assumed to lead to stochastic acceleration of
energetic electrons which are responsible for the synchrotron emission
from the radio to the X-ray band 
\citep{Scott_ApJ_1975_197,Dickel_AA_1979_75,Coswick_MNRAS_1984_207,Jones_ApJ_1994_432,Atoyan_AA_2000_354}.
This gives also rise to an attendant nonthermal Bremsstrahlung (NB) and Inverse
Compton (IC) component extending to the high energy $\gamma$-rays. In this
phenomenological scenario the large scale SNR blast wave does not have to
play an important role for electron energization, but it may still be the
main source for the nonthermal nuclear particle component 
\citep{Atoyan_AA_2000_355}.
An inventory of the radio to high energy \gr emission along these
lines has been made phenomenologically by 
\citet{Atoyan_AA_2000_354,Atoyan_AA_2000_355},
including detailed fits to the observed synchrotron spectrum.

Alternatively one can to lowest order ignore the role of small-scale
inhomogeneities for the production of the {\it very high energy}
nonthermal particle component. At least as its carriers they are in
difficulty on the argument that high energy particles (ultrarelativistic
electrons and nuclear particles) require a large acceleration/propagation
volume given by spatial scales $L \sim \lambda_{\mathrm{mfp}}(p)\times c/u$. Here
$u$ denotes a characteristic nonrelativistic speed of mass motions, $c$ is
the speed of light, and $\lambda_{\mathrm{mfp}}(p)$ is the scattering mean free
path that increases with momentum $p$. For strong scattering
$\lambda_{\mathrm{mfp}} = \mathrm{O}(r_{\mathrm{g}})$, where $r_{\mathrm{g}} \propto p$ is the particle
gyro radius. At least for the highest energy particles $L$ amounts to a
few percent of the remnant radius. Ignoring the dynamical effects
underlying the small-scale emission features, the dominant remaining
large-scale entropy generator is the SNR blast wave. It has then to be
investigated which of the observed nonthermal features can be explained by
shock accelerated particles.

In this paper we shall apply the nonlinear kinetic model for diffusive
shock acceleration in SNRs 
\citep{Berezhko_Jetph_1996_82,Berezhko_Astropart_1997_7,Berezhko_Astropart_2000_14,Berezhko_AA_2000_357}
to the specific case of Cas\,A. This spherically symmetric model
should describe the global properties of the nonthermal emission
especially at high energies, even though small-scale processes may indeed
introduce some modifications at low particle energies, contributing
especially to the radio flux. The calculation of the nonthermal particle
populations is a central part of the theory which also determines the
spherically symmetric portion of the overall gas dynamics in the remnant
selfconsistently.

In fact, we shall attempt to investigate whether the global properties of
Cas\,A are consistent with the idea that the SN blast wave is the {\it
main} source of energetic particles and in particular, whether the
observed \gr emission is consistent with a {\it hadronic} origin.


\section{Model}

To describe the circumstellar medium we use the specific model of
\citet{Borkowski_ApJ_1996_466}.
Accordingly, part of the slow red supergiant wind
of the SN progenitor has been swept up into a dense shell by a
fast stellar wind during the final blue supergiant (probably Wolf-Rayet)
phase of the progenitor star. Therefore the inner circumstellar medium 
consists of three
zones: a tenuous wind-blown bubble, a dense shell, and a freely expanding
red supergiant wind. The outer circumstellar regions that were generated during
the main sequence phase play no role here.

We describe the profile of gas number density $N_{\mathrm{g}}=\rho/m_{\mathrm{p}}$ in the 
analytic form
\begin{equation}
N_{\mathrm{g}}=\frac{N_{\mathrm{bsh}}+N_{\mathrm{w}}}{2}+\frac{N_{\mathrm{w}}-N_{\mathrm{bsh}}}{2} \tanh\left(
\frac{r-R_2}{l}\right),
\end{equation}
where
\begin{equation}
N_{\mathrm{bsh}}=\frac{N_{\mathrm{b}}+N_{\mathrm{sh}}}{2}+\frac{N_{\mathrm{sh}}-N_{\mathrm{b}}}{2} \tanh\left(
\frac{r-R_1}{l}\right),
\end{equation}
and
\begin{equation}
N_{\mathrm{w}}=N_{\mathrm{w2}}(R_2/r)^2
\end{equation}
is the gas number density of the free red giant wind (region $r>R_2$),
$N_{\mathrm{b}}$ and $N_{\mathrm{sh}}$ correspond to the hot bubble (region $r<R_1$) and shell
(region $R_1<r<R_2$) respectively, $\rho$ is the gas density, $m_{\mathrm{p}}$ is the
proton mass. This interpolation formula provides a smooth transition
between the above three zones on the scale $l$ which is taken small
enough, $l\ll R_1$, that its concrete value does not influence the final
results.

We use the same type of formula (1) and (2) for the magnetic field
profile $B_0(r)$
with $B_{\mathrm{b}}$, $B_{\mathrm{sh}}$ and 
\begin{equation}
B_{\mathrm{w}}=B_{\mathrm{w2}} R_2/r
\end{equation}
for the bubble, shell and wind regions
respectively.

The SN explosion ejects an expanding amount of matter with
energy $E_{\mathrm{sn}}$ and mass $M_{\mathrm{ej}}$. During an initial period the ejecta
have a wide distribution in radial velocity $v$. The fastest part of these
ejecta is described by a power law $\mathrm{d}M_{\mathrm{ej}}/\mathrm{d}v \propto v^{2-k}$
\citep{Jones_ApJ_1981_249}. 
The interaction of the ejecta with the circumstellar medium 
creates a strong
shock which will diffusively accelerate particles.

The acceleration model consists in a selfconsistent solution of the CR
transport equation together with the gas dynamic equations in spherical
symmetry 
\citep{Berezhko_Jetph_1996_82,Berezhko_Astropart_1997_7},
in an extension of this model to the case of a nonuniform circumstellar medium
\citep{Berezhko_AA_2000_357}.
 
The CR diffusion coefficient is taken at the Bohm limit
\begin{equation}
\kappa (p)=\kappa (mc) (p/mc),
\label{EQ:Bohm}
\end{equation}
where $\kappa(mc)=mc^3/(3eB)$; $e$, $m$ and $p$ are the particle charge,
mass and momentum, respectively; $B$ is the magnetic field strength;
and $c$ is the speed of light.

In the downstream region the magnetic field is assumed to be frozen into
the gas and is described by the equation
\begin{equation}
\frac{\partial \vec{B}} {\partial t}= \nabla \times (\vec{w} \times 
\vec{B}),
\end{equation}
where $\vec{w}$ is the gas mass speed, directed everywhere radially in our
spherical model. 
The boundary condition for this equation is the relation between the postshock
magnetic field $B_{2}$ and the upstream field at the shock position
$B_{\mathrm{s}}=B_{0}(R_{\mathrm{s}})$. The tangential field component $B_{\perp}$ 
is amplified at the shock front because of compression, so that its postshock value
is $B_{2\perp}=\sigma B_{0\perp}$, where $\sigma$ is the shock compression ratio.
Therefore we have $B_2=\sigma_\mathrm{B} B_0$, where 
$\sigma_\mathrm{B} = \sqrt{\left(\sigma^2-1\right) \sin^2\phi +1}$,
$\phi$ being the angle between the upstream magnetic field $\vec{B}_0(R_\mathrm{s})$
and the shock normal.
For simplicity we suggest that the downstream magnetic
field is purely tangential and use $\sigma_{\mathrm{B}}=\sigma$.

The number of suprathermal protons injected into the acceleration process
is described by a dimensionless injection parameter $\eta$ which is a
fixed fraction of the ISM particles entering the shock front. For
simplicity it is assumed that the injected particles have a velocity four
times higher than the postshock sound speed. Unfortunately there is no
complete selfconsistent theory of a collisionless shock transition, which
can predict the value of the injection rate and its dependence on the
shock parameters. For the case of a purely parallel shock hybrid
simulations predict quite a high ion injection 
\citep[e.g.][]{Scholer_ApJ_1992_395,Bennet_JGR_1995_100}
which corresponds to the value $\eta \sim 10^{-2}$
of our injection parameter. Such a high injection is consistent with
analytical theory 
\citep{Malkov_AA_1995_300,Malkov_AdvSpaceRes_1996_21,Malkov_PhysRevE_1998_58}
and confirmed by measurements near the Earth's bow shock 
\citep{Trattner_JGR_1994_99}.
We note however that in our spherically symmetric model these results can
only be used with some important modification. The circumstellar magnetic
field may be assumed to be strongly perturbed at least in the shell
region. Efficient particle injection takes place only on those portions of
the shock surface which are currently locally quasiparallel. On other
parts of the shock where it is more and more oblique, the magnetic
field essentially suppresses the leakage of suprathermal particles from
the downstream region back upstream 
\citep{Ellison_ApJ_1995_453,Malkov_AA_1995_300}.
Therefore the mean injection rate and subsequent CR production efficiency,
properly averaged over the
entire shock surface, is expected to be considerably lower compared with
the case of a purely parallel shock. 
Due to this fact the number of accelerated CRs, calculated within our
spherically symmetric model, has to be corrected (decreased) by some
renormalisation factor $f_{\mathrm{re}}<1$, because of the lack of efficient 
injection/acceleration on the part $1-f_{\mathrm{re}}$ of the actual shock surface.

We assume that electrons are also injected into the diffusive shock
acceleration process still at nonrelativistic energies below 
$m_{\mathrm{e}} c^2$. 
Since the electron injection mechanism is not very well known 
\citep[e.g.][]{Malkov_RepProgPhys_2001_64},
for simplicity we consider their
acceleration starting from the same momentum as protons. At relativistic
proton energies they have exactly the same dynamics as the protons.
Therefore, neglecting synchrotron losses, their distribution function at
any given time has the form 
\begin{equation} 
  f_{\mathrm{e}}(p)= K_{\mathrm{ep}} f(p) 
  \label{EQ:electronprotonratio} 
\end{equation} 
for relativistic proton energies, with some constant factor
$K_{\mathrm{ep}}$ which is about $10^{-2}$ for the average CRs in the
Galaxy. We take it here as a parameter
and use relation (\ref{EQ:electronprotonratio}) only at the injection momentum in order to determine
the number of injected/accelerated electrons for a given proton injection rate.
At all other momenta the electron distribution function is 
found by solving the transport equation.

In fact the calculated electron distribution function $f_{\mathrm{e}}(p)$ deviates 
from expression (\ref{EQ:electronprotonratio}) only at sufficiently
large momenta because of synchrotron losses, which are
taken into account by supplementing the ordinary diffusive transport
equation with a loss term: 
\begin{equation} 
  {\partial f_{\mathrm{e}} \over \partial t} 
  = 
  \nabla \kappa \nabla f_{\mathrm{e}} -\vec{w} \nabla f_{\mathrm{e}} 
  + \frac{\nabla \vec{w}}{3} p \frac{\partial f_{\mathrm{e}}}{\partial p}
  + \frac{1}{p^2} \frac{\partial}{\partial p} 
    \left( \frac{p^3}{\tau_{\mathrm{l}}} f_{\mathrm{e}} \right), 
\end{equation} 
where the first three terms on the right hand side of this equation
describe diffusion, convection due to the mass velocity $\vec{w}$ of the
gas, and adiabatic effects, respectively. The synchrotron loss time in the
third term is determined by the expression 
\citep[e.g.][]{Berezinskii_Astrophysics_1990}
\begin{equation} 
  \tau_{\mathrm{l}} = \left( \frac{4\,r_0^2 B^2p}{9\,m_{\mathrm{e}}^2c^2}\right)^{-1}, 
  \label{EQ:synchrotronlosstime}
\end{equation} 
where $m_{\mathrm{e}}$ is the electron mass and $r_0$ the classical electron radius.

The solution of the dynamic equations at each instant of time yields the
CR spectrum and the spatial distributions of CRs and gas.  This allows us
to calculate the expected flux $F_{\gamma}^\mathrm{\pi}(\epsilon_{
\gamma})$ of $\gamma$-rays from $\pi^0$-decay because of hadronic (p-p)  
collisions of CRs with the gas nuclei. Following the work of 
\citet{Dermer_AA_1986_157}
and its later improvement by 
\citet{Naito_JPhysG_1994_20}
we use here the isobar model at the proton kinetic
energies $\epsilon_{\mathrm{k}}<3\,\mathrm{GeV}$ and the scaling model at
$\epsilon_{\mathrm{k}}>7\,\mathrm{GeV}$
with a linear connection between $3$ and $7\,\mathrm{GeV}$. This model agrees very well
with the simpler approach, introduced by 
\citet{Drury_AA_1994_287} \citep[see also][]{Berezhko_Astropart_1997_7,Berezhko_Astropart_2000_14,Berezhko_ProcICRC_1999_4}
at high energies $\epsilon_{\gamma}>0.1\,\mathrm{GeV}$, except in the
cutoff region, where the scaling model yields a significantly smoother
turnover of the $\gamma$-ray spectrum at somewhat lower energies.

The choice of $K_{\mathrm{ep}}$ allows us to determine in addition the electron
distribution function and to calculate the associated emission
\citep[for details, see][ in a recent analysis of SN\,1006]{Berezhko_AA_2002_395}.
We calculate here the IC radiation taking into account as target photon fields 
-- besides the cosmic microwave background --
the infrared field with a mean photon energy of $\epsilon_\mathrm{ph}=0.01\,\mathrm{eV}$ 
and an energy density of $0.2\,\mathrm{eV}/\mathrm{cm}^3$,
and the optical field with a mean photon energy of 
$\epsilon_\mathrm{ph}=1.5\,\mathrm{eV}$
and an energy density of $0.5\,\mathrm{eV}/\mathrm{cm}^3$
\citep[e.g.][]{Drury_AA_1994_287,Gaisser_ApJ_1998_492}.

\section{Results and Discussion}

\subsection{Supernova dynamics}

\begin{figure} 
\centering 
\includegraphics[width=7.5cm]{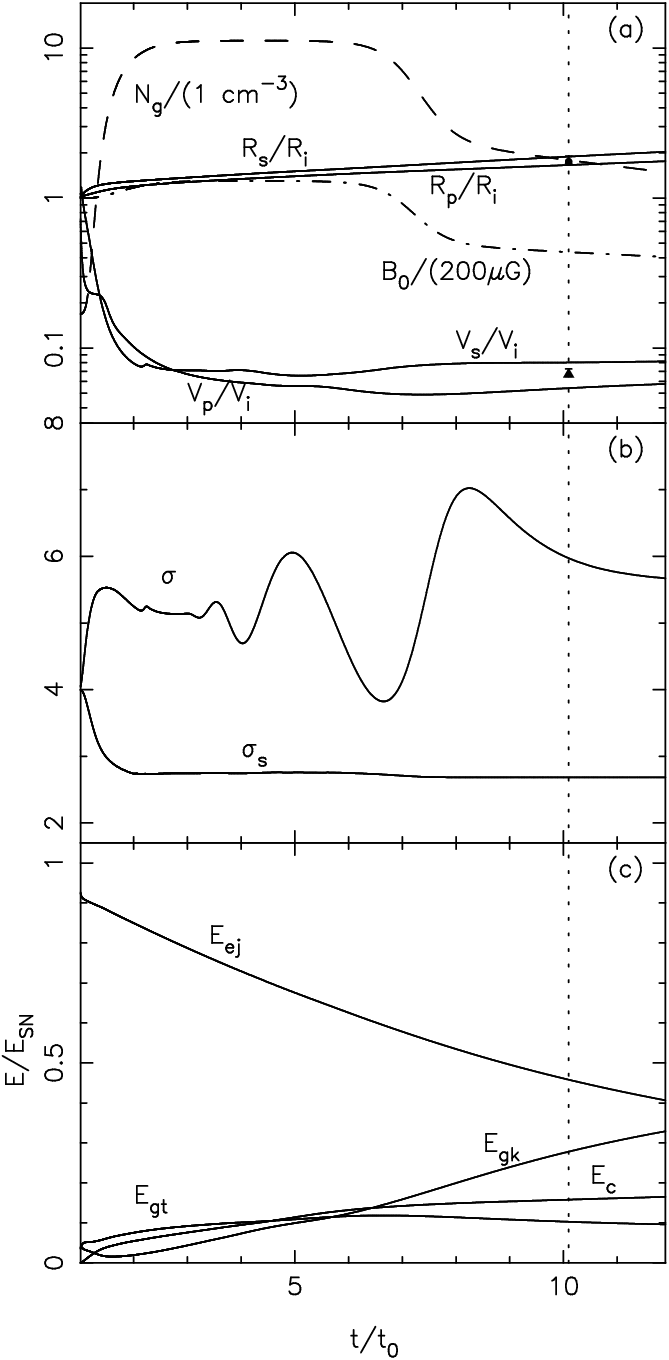}
\caption{Shock (contact discontinuity) radius $R_{\mathrm{s}}$ ($R_{\mathrm{p}}$) and shock
(contact discontinuity) speed $V_{\mathrm{s}}$ ($V_{\mathrm{p}}$) ({\bf a}); total shock
($\sigma$) and subshock ($\sigma_{\mathrm{s}}$) compression ratios ({\bf b}); ejecta
($E_{\mathrm{ej}}$), CR ($E_{\mathrm{c}}$), gas thermal ($E_{\mathrm{gt}}$) and gas kinetic ($E_{\mathrm{gk}}$)
energies ({\bf c}) as a function of time. The vertical dotted line
corresponds to the current evolutionary stage. The circumstellar gas
number density $N_{\mathrm{g}}(R_{\mathrm{s}})$ and magnetic field $B_0(R_{\mathrm{s}})$ profiles are shown
in ({\bf a}) by dashed and dash-dotted lines, respectively. Scale values are
$R_{\mathrm{i}}=1\,\mathrm{pc}$, $V_{\mathrm{i}} = 30\,000\,\mathrm{km/s}$, $t_0=31.7\,\mathrm{years}$.
The observed size 
\citep{Reed_ApJ_1995_440}
and speed 
\citep{Andersen_ApJ_1995_441}
of the shock are shown
by the full circle and triangle, respectively, in ({\bf a}).}
\label{Figure1}
\end{figure}

In the current approach, we have reduced the circumstellar density profile
everywhere by a factor of 1.8 compared with what was derived by \citet
{Borkowski_ApJ_1996_466}. This corresponds to the parameter values:
\[N_{\mathrm{b}} = 0.17\,\mathrm{cm}^{-3},\hspace{0.2cm}
  N_{\mathrm{sh}} = 11\,\mathrm{cm}^{-3},\hspace{0.2cm}
  N_{\mathrm{w2}} = 2\,\mathrm{cm}^{-3},
\]
\[
  R_1 = 3.8 \times 10^{18}\mathrm{cm},\hspace{0.4cm}
  R_2 = 4.94 \times 10^{18}\mathrm{cm}.
\]
Together with the renormalization factor $f_{\mathrm{re}}$ which is
discussed later, the lower circumstellar density
is the main reason for the significant reduction of
the expected $\gamma$-ray flux (see below) compared to our previous
calculations which did not take into account $f_{\mathrm{re}}$ and were based
on a swept-up mass value of $8M_{\odot}$ \citep{Berezhko_ProcICRC_2001_2}.
The density reduction is close to that suggested by the X-ray analysis of
the BeppoSAX data \citep{Favata_AA_1997_324}, which yielded  $5M_{\odot}$
instead of $8M_{\odot}$ \citep{Vink_AA_1996_307,Borkowski_ApJ_1996_466}
for the swept-up mass.


In order to reproduce the observed shock size $R_{\mathrm{s}}$ and its expansion rate
$V_{\mathrm{s}}$ the following SN parameters are used: explosion energy
$E_{\mathrm{sn}} = 4 \times 10^{50}\mathrm{erg}$, ejecta mass $M_{\mathrm{ej}} = 2M_{\odot}$, and
power-law index $k=6$ for the ejecta velocity distribution
\citep[cf.][]{Berezhko_AA_2002_395}.
As distance to Cas\,A we adopt $d=3.4\,\mathrm{kpc}$ 
\citep[cf.][]{Reed_ApJ_1995_440}.

The results of our calculations together with the experimental data are
shown in Fig.\,\ref{Figure1}--\ref{Figure4}. In Fig.\,\ref{Figure1}a
we also show the profiles of the gas number
density $N_{\mathrm{g}}(R_{\mathrm{s}})$ and magnetic field $B_0(R_{\mathrm{s}})$, upstream of the shock
front that is at position $R_{\mathrm{s}}(t)$.

As can be seen in Fig.\,\ref{Figure1}a, after the shock has entered the shell region its
speed drops during the initial 60 years by more than a factor of ten and
then remains almost constant up to the current epoch.

At the current epoch $t \approx 320\,\mathrm{yr}$ the calculations reasonably
reproduce the observed size 
\citep{Reed_ApJ_1995_440}
and expansion rate
\citep{Andersen_ApJ_1995_441}
of the remnant.

A proton injection rate $\eta = 2.5 \times 10^{-3}$ is used in order to
provide the nonlinear shock modification required to reproduce quite a
steep radio emission spectrum (see below). According to Fig.\,\ref{Figure1}b the shock
is indeed strongly modified by the CR backreaction: the total shock
compression ratio $\sigma \approx 5.5$ exceeds the classical value $4$,
whereas the subshock compression ratio is considerably smaller, 
$\sigma_{\mathrm{s}} \approx 2.7$.

\subsection{The accelerated CR spectrum}

According to Fig.\,\ref{Figure1}c about 15\% of the explosion energy has 
been transformed into CRs at the current stage, that is 
$E_{\mathrm{c}} \approx 6 \times 10^{49}\mathrm{erg}$.

\begin{figure}
  \centering 
  \includegraphics[width=7.5cm]{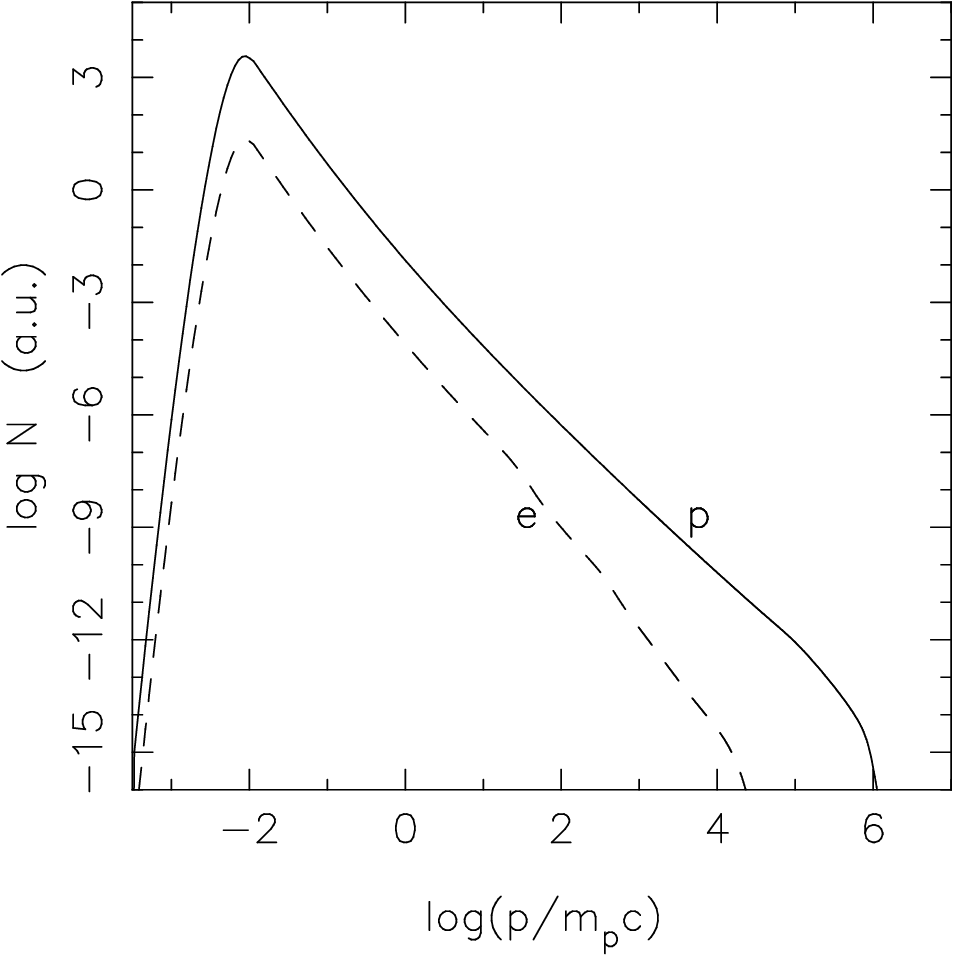}
  \caption{The overall CR proton ({\it solid line}) and electron ({\it dashed
  line}) spectra as function of momentum.} 
  \label{Figure2}
\end{figure}
The overall momentum spectrum of accelerated protons 
\begin{equation}
  N(p,t) = 16 \pi^2p^2 \int_0^{\infty}\mathrm{d}r\,r^2 f(r,p,t), 
\end{equation} 
presented in Fig.\,\ref{Figure2}, turns over at about $p = 10^5m_{\mathrm{p}}c$ and
extends almost up to $p_{\mathrm{max}} = 10^6 m_{\mathrm{p}}c$ because of the extremely
high magnetic field strength used: $B_{\mathrm{sh}} = 200\,\mu\mathrm{G}$,
$B_{\mathrm{w2}} = 100\,\mu\mathrm{G}$.  The value $p_{\mathrm{max}}$ is limited by
geometrical factors which are the finite size and speed of the shock, its
deceleration and the adiabatic cooling effect in the downstream region
\citep{Berezhko_Astropart_1996_5}. 
If we approximate the overall proton spectrum by a
power-law $N\propto p^{-\gamma}$ then the power-law index slowly decreases
from $\gamma\approx 3$ at $p \lsim m_{\mathrm{p}}c$ to $\gamma\approx 1.9$
at the highest momenta $10^2 m_{\mathrm{p}} \lsim p \lsim 10^5m_{\mathrm{p}}c$
due to the strong shock modification.

The shape of the overall electron spectrum deviates from proportionality
to the proton spectrum $N_{\mathrm{e}}(p) \propto N(p)$ at high momenta 
$p > p_{\mathrm{l}} \approx 30\,m_{\mathrm{p}}c$, because of the synchrotron losses in the downstream
region where the magnetic field 
$B_{\mathrm{d}}\approx B_2$ is of the order of $1\,\mathrm{mG}$.
According to expression (\ref{EQ:synchrotronlosstime}) the synchrotron losses become important for
electron momenta greater than
\begin{equation} 
  \frac{p_{\mathrm{l}}}{m_{\mathrm{p}}c} \approx 1.3
  \left(\frac{10^8\,\mathrm{yr}}{t}\right)
  \left(\frac{10\,\mu\mathrm{G}}{B_{\mathrm{d}}}\right)^2. 
\end{equation}

Substituting the SN age $t = 320\,\mathrm{yr}$ into this expression we have
$p_{\mathrm{l}} \approx 30\,m_{\mathrm{p}}c$, in good agreement with the numerical
results
(Fig.\,\ref{Figure2}).

The shock constantly produces the electron spectrum $f_{\mathrm{e}} \propto p^{-q}$
with $q \approx 4$
up to the maximum momentum $p_{\mathrm{max}}^{\mathrm{e}}$, which is much larger than
$p_{\mathrm{l}}$. Therefore, within the momentum range $p_{\mathrm{l}}$ to $p_{\mathrm{max}}^{\mathrm{e}}$
due to the synchrotron losses, the electron spectrum is $f_{\mathrm{e}} \propto p^{-5}$
and the corresponding overall electron spectrum is $N_{\mathrm{e}} \propto p^{-3}$.

The maximum electron momentum can be estimated by equating the 
synchrotron loss time (\ref{EQ:synchrotronlosstime}) and the acceleration time
\begin{equation}
  \tau_{\mathrm{a}} = \frac{3}{\Delta u} \left(
  \frac{\kappa_1}{u_1}+\frac{\kappa_2}{u_2},
  \right)
\end{equation}
where $u_1 = V_{\mathrm{s}}$ and $u_2 = u_1/\sigma$ are the upstream and downstream gas
velocities with respect to the shock front, and $\Delta u = u_1-u_2$. In the
general case the downstream magnetic field equals $B_2 = \sigma_{\mathrm{B}} B_0$
and correspondingly
\begin{eqnarray}
  \frac{p_{\mathrm{max}}^e}{m_{\mathrm{p}}c}
  &=& 6.7\times 10^4 \left(\frac{V_{\mathrm{s}}}{10^3\,\mathrm{km/s}}\right) 
  \nonumber \\
  &\times& \sqrt{\frac{(\sigma-1)}{\sigma (1+\sigma_{\mathrm{B}} \sigma)}
  \left(\frac{10\,\mu\mathrm{G}}{B_0}\right)}. 
\end{eqnarray}
The main part of the electrons with the highest energies 
$\epsilon\gsim 10\,\mathrm{TeV}$ is produced during shock propagation through the shell. At this
stage $V_{\mathrm{s}} \approx 2500\,\mathrm{km/s}$ which leads to a maximum electron momentum
$p_{\mathrm{max}}^{\mathrm{e}} \approx 10^4m_{\mathrm{p}}c$ 
(for simplicity we use $\sigma_{\mathrm{B}} = \sigma$ below), 
in agreement with the numerical results (Fig.\,\ref{Figure2}).

According to the results from the next section,
the energy content of the electrons at the current epoch is 
$E_{\mathrm{c}}^{\mathrm{e}} = K_{\mathrm{ep}}E_{\mathrm{c}} = 2.4 \times 10^{47}\mathrm{erg}$.

\subsection{The synchrotron emission from Cas\,A}

\begin{figure}
  \centering
  \includegraphics[width=7.5cm]{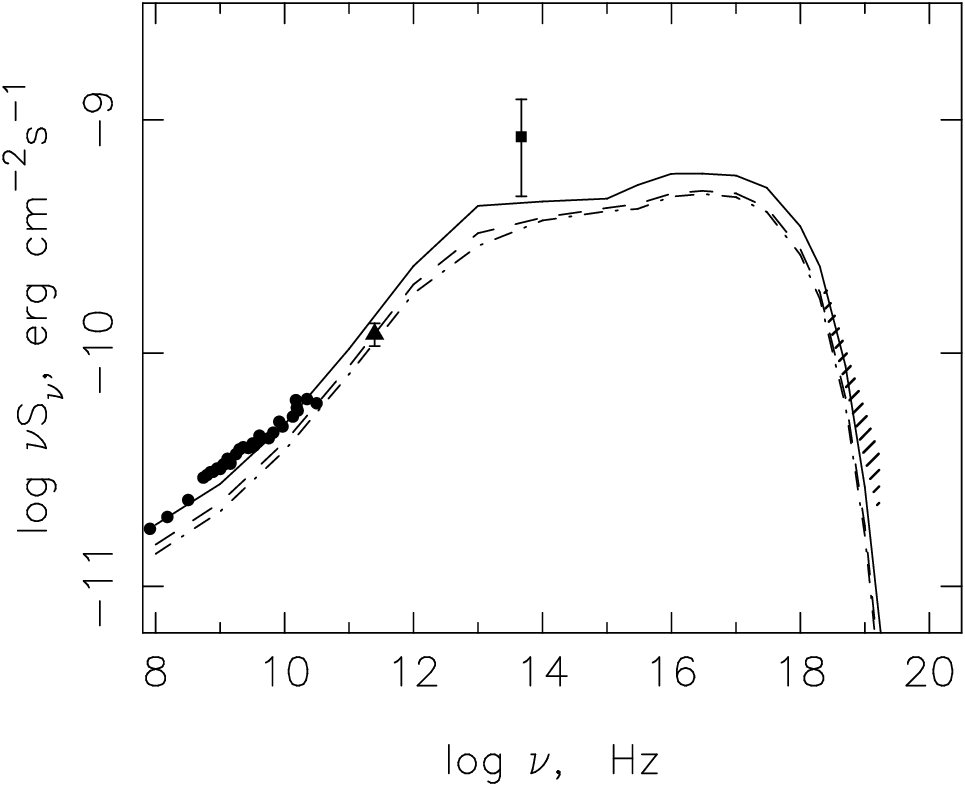}
  \caption{
  Overall synchrotron spectral energy distributions as a function of
  frequency, calculated for the epochs 1970 (solid curve), 2002 (dashed curve), and 2022 (dashed-dotted
  curve). The radio-emission above $100\,\mathrm{MHz}$
  \citep{Baars_AA_1977_61},
  the data at $1.2\,\mathrm{mm}$ (triangle) from 
  \citet{Mezger_AA_1986_167}
  and $6\,\mu\mathrm{m}$ (square) from 
  \citet{Tuffs_ProcISO_1997},
  as well as the hard X-ray spectrum 
  \citep{Allen_ApJ_1997_487}
  are presented.
  Due to the non-contemporaneous observations, the radio data should correspond to the solid
  curve, whereas the X-ray data must be compared with the dashed curve. In the radio range
  theory and experiment agree fairly well. To achieve agreement also in the X-ray regime, a
  small increase of the overall electron cutoff momentum $p_{\mathrm{max}}$ of roughly 10\%
  would be required (see text for details).
  }
  \label{Figure3}
\end{figure}

The parameter $K_{\mathrm{ep}} = 4 \times 10^{-3}$ gives reasonable agreement 
between
the calculated and the measured synchrotron emission in the radio- and X-ray
ranges, as one can see from Fig.\,\ref{Figure3}. The calculated synchrotron
spectrum at the epoch 1970 (solid curve), which corresponds to that of the radio measurements of
\citet{Baars_AA_1977_61}, is compared with the experimental data. 
Also a calculated curve for the present epoch (dashed line), corresponding to the X-ray 
measurements, and a prediction for the epoch 2022 (dashed-dotted line) are given.
The strong
shock modification leads to a steep spectrum $N_{\mathrm{e}} \propto p^{-2.6}$ of
sub-GeV accelerated particles (Fig.\,\ref{Figure2}). As result of our choice of a
large magnetic field ($B_{\mathrm{sh}} = 200\,\mu\mathrm{G}$,
$B_{\mathrm{w2}} = 100\,\mu\mathrm{G}$) the calculation fits the radio data $S_{\nu} \propto
\nu^{-\alpha}$, $\alpha \approx 0.8$, rather well. In fact, the electron
spectrum has a concave shape at $p < p_{\mathrm{l}} \approx 30\,m_{\mathrm{p}}c$. This
leads to a flattening of the synchrotron spectrum $S_{\nu}(\nu)$ at $\nu=
10$ to $100\,\mathrm{GHz}$, consistent with the experiment.  Our calculated
synchrotron flux fits even marginally the 
\citet{Mezger_AA_1986_167}
measurements at $1.2\,\mathrm{mm}$ and the infrared emission measured at $6\,\mu\mathrm{m}$ by
\citet{Tuffs_ProcISO_1997},
even though, as one can see from Fig.\,\ref{Figure3}, these two points are above
the pure power-law extrapolation $S_{\nu}\propto \nu^{-0.77}$. It is
possible that the far infrared energy flux has a significant thermal 
component, cf.
\citet{Braun_AA_1987_171}, \citet{Tuffs_ProcISO_1997} and \citet{Vink_Phd_1999}.
The steepening of the electron spectrum at $p > p_{\mathrm{l}}$ caused by
synchrotron losses leads to a flat spectral energy distribution $\nu S_{\nu}$
that connects the nonthermal mid-infrared with the X-ray band
(Fig.\,\ref{Figure3}).

It is important to note that the complicated shape of the electron spectrum
is naturally produced by the model. As far as the overall nonthermal
spectrum is concerned, there is no strong need to add other sources of
energetic particles. The bright `radio ring' is in any case explainable by
the swept-up red supergiant wind shell which is contained in our model. It
is another question whether individual `radio knots' are local
acceleration regions or rather just magnetic field enhancements bathed in
a pervasive particle background. For the nonthermal X-ray and TeV \gr
emission they appear too small to locally accelerate the radiating
multi-TeV particles in a stochastic manner. For a given frequency, a radio
knot that consists of a local B-field enhancement will be illuminated by
lower energy electrons than the ambient spatial regions. The strong
nonlinear modification of the SN shock that generates such particles then
ensures that these lower energy electrons have a steeper spectrum -- just
what distinguishes these knots form the average emission. Whether this
effect is quantitatively sufficient, is a difficult question on the
characteristics of the turbulent structure of the shocked red supergiant
wind shell. However, it might go a long way to explain the 20\% contribution 
\citep{Tuffs_MNRAS_1986_219}
of the radio knots to the total radio emission.
Potential $2^{\mathrm{nd}}$ order Fermi acceleration in the turbulent shell, as
proposed by 
\citet{Scott_ApJ_1975_197},
cannot be ruled out. However, there
is no indication for its action either. For diffusive shock acceleration at the
blast wave, on the other hand, we know from theory that it can have the
required high efficiency. And it can do it all. No other process needs to
be invoked.

The shock modification can only be produced by the backreaction of the
proton CR component. Therefore one can consider the extremely steep Cas\,A
spectrum in the radio range as indirect evidence of efficient nuclear CR
production. The other important physical factor is the high magnetic field
value which leads to substantial synchrotron losses of electrons with
energies $\epsilon_{\mathrm{e}} > 30\,\mathrm{GeV}$: a magnetic field strength $B_{\mathrm{sh}} = 200 \mu\mathrm{G}$
is neded to reproduce the observed radio and X-ray synchrotron fluxes.
Since we assume that the postshock field $B_2 = \sigma B_{\mathrm{s}}$, the downstream
magnetic field amounts to about $B_{\mathrm{d}} \approx B_2 \approx 1\,\mathrm{mG}$ in the shell
(Fig.\,\ref{Figure1}a). This is roughly consistent with previous estimates 
\citep[e.g.][]{Atoyan_AA_2000_354}.

Such a large magnetic field $B_0(r)$ considerably exceeds the value for
a red supergiant wind, assuming that its typical surface field at a radius of about $3\times
10^{13}$~cm is about 1 G. We believe that the large field strength in the
shell is likely either due to turbulent amplification of the red supergiant
wind field by the shell formation in the final Wolf-Rayet phase or due to
considerable field amplification near the shock by CR streaming
\citep{Lucek_MNRAS_2000_314}. If the second factor is relevant, the expected
amplified field goes like $B_0(R_{\mathrm{s}}) \propto
\sqrt{N_g(R_{\mathrm{s}})}$ \citep{Bell_MNRAS_2001_321}. Our field
$B_0(R_{\mathrm{s}})$ is distributed according to this relation.
The deduced field amplification is still well within the upper bound set by 
the overall energy and momentum balance relations \citep{Voelk_AA_2002_396}.

Our results show that even at the current epoch (when the SN shock
propagates through the free red supergiant
wind) the observed synchrotron emission is
still determined by the electrons accelerated during the shock propagation
through the shell.  Therefore the magnetic field $B_0(r)$ in the wind zone
is not a very relevant parameter for the fit to the observations.
 
According to Eq.\,(\ref{EQ:Bohm}), we note also
that the exponential cutoff of the overall synchrotron spectrum 
at $\nu_{\mathrm{max}}\sim 10^{18}\mathrm{Hz}$ is roughly independent of the magnetic field,
as pointed out by \citet{Aharonian_AA_1999_351} for the case of a uniform field.
Indeed the emission with frequency $\nu \sim \nu_{\mathrm{max}}$ is produced by
electrons with momenta $p \sim p_{\mathrm{max}}^{\mathrm{e}}$ from their cutoff region. 
Since the maximum power of the synchrotron emission of electrons with momentum
$p$ is emitted at frequency $\nu \propto \sigma_{B} B_{0} p^2$ 
\citep[e.g.][]{Berezinskii_Astrophysics_1990},
and taking into account that $p^{\mathrm{e}}_{\mathrm{max}} \propto V_{\mathrm{s}} /
\left(\sigma_{B} \sqrt{B_0}\right)$ we
conclude that $\nu_{\mathrm{max}} \propto V_{\mathrm{s}}^{2}/\sigma_{B}$, independent of $B$.
A roughly 20\% increase in $\nu_{\mathrm{max}}$ would be required to bring the overall
X-ray data into close agreement with the contemporaneous model calculations (dashed curve),
while leaving the low-frequency ranges unchanged at all epoches (the low-energy electron
spectrum is independent of the cutoff energy). Whereas the spectral hardening beyond the radio
range is a basic feature of nonlinear shock acceleration, this is not true for the precise
value of the overall high-energy cutoff. This cutoff depends on the detailed characteristics
of the scattering wave field at long wavelengths. With the above approximation of the
diffusion coefficient by the Bohm limit, we see that a 20\% increase in the characteristic
value of $V_{\mathrm{s}}^{2}/\sigma_{B}$ is well within the variations of these quantities
over the shell region (Fig.\ref{Figure3}). Thus we can consider the agreement between
theoretical results and observations as remarkably good.

\begin{figure}
  \centering
  \includegraphics[width=7.5cm]{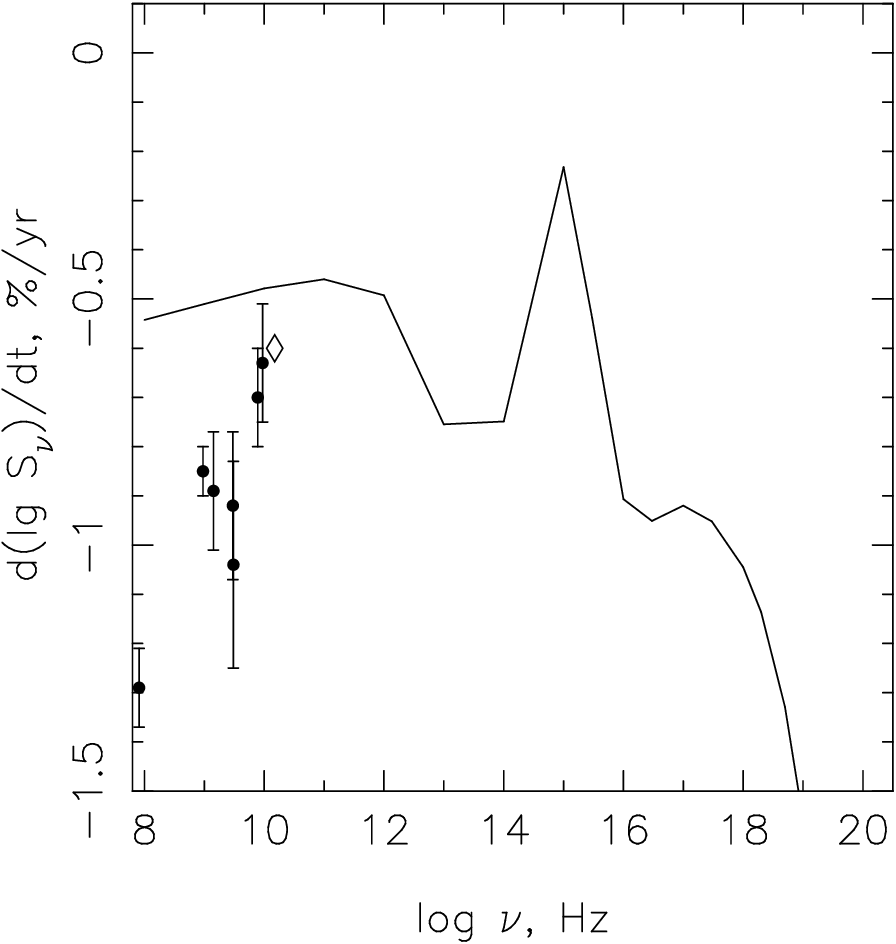}
  \caption{The rate of secular decline of the total flux of synchrotron
    radiation. The diamond shows the recent result of 
    \citet{Sullivan_MNRAS_1999_303}
    at $15\,\mathrm{GHz}$. All other data points are from
    \citet{Rees_MNRAS_1990_243}.
    }
  \label{Figure4}
\end{figure}

\subsection{The secular decline of the radio flux}

Important information about the circumstellar parameters comes from 
the measured secular decline of the radio flux which is presented in
Fig.\,\ref{Figure4}. 
In the case of a very young SNR the population of CRs would be expected to be an
increasing function of time, if the remnant was expanding into a homogeneous medium
with constant density and magnetic field.
For Cas\,A, even the existence of the decline of the radio flux itself
is a strong indication that the SN shock at 
the current evolutionary stage propagates through a region
of monotonically decreasing magnetic field strength $B_0(R_{\mathrm{s}})$ and 
density
$\rho_0(R_{\mathrm{s}})$. The most natural assumption, proposed by 
\citet{Borkowski_ApJ_1996_466},
identifies this region with the wind region. If we were to 
suppose that the CR 
population and the radio emission from its electron component are
dominated by particles accelerated in this free wind region, the 
synchrotron flux would scale as
\begin{equation}
  S_{\nu} \propto B_0(R_{\mathrm{s}})^{(\gamma +1)/2}N_0(R_{\mathrm{s}}),
\end{equation}
where we represented the spectrum of radio emitting electrons in 
power-law form
\begin{equation}
  N_{\mathrm{e}}(p) = N_0(p/p_0)^{-\gamma}.
\end{equation}
In the wind region we have $B_0(R_{\mathrm{s}}) \propto R_{\mathrm{s}}^{-1}$, the number of
accelerated electrons scales as the amount of swept-up mass  $N_0 \propto R_{\mathrm{s}}$,
and therefore 
\begin{equation}
  S_{\nu} \propto R_{\mathrm{s}}^{-(\gamma-1)/2}.
\end{equation}
In our case $V_{\mathrm{s}} \approx 2400\,\mathrm{km/s}$, $R_{\mathrm{s}} \approx 2\,\mathrm{pc}$ 
and $\gamma \approx 2.6$. For the secular flux decline
\begin{equation}
  \frac{\mathrm{d}\lg S_{\nu}}{\mathrm{d}t} = -\frac{\gamma-1}{2}
  \frac{V_{\mathrm{s}}}{R_{\mathrm{s}}}
\end{equation}
we would therefore have $\mathrm{d}\lg S_{\nu}/\mathrm{d}t \approx - 0.1$\%$/\mathrm{yr}$.
Its absolute value is
considerably smaller than the observed one (Fig.\,\ref{Figure4}).

The explanation for the large observed value of the secular decline of
the radio flux comes from the idea that the radio emission is still
dominated by the contribution of electrons accelerated at the previous 
stage during shock propagation through the dense shell. In this
case the electrons undergo adiabatic cooling because of the expansion of
the downstream region, and one can rewrite relation (14) in the form
\begin{equation}
  S_{\nu} \propto B_{\mathrm{d}}^{(\gamma +1)/2}a^{-(\gamma+2)},
\end{equation}
where $a = (\rho_{\mathrm{i}}/\rho)^{1/3}$ is the adiabatic factor which describes 
the particle adiabatic cooling, $\rho$ is the
gas density in the region where electrons are confined, and $\rho_{\mathrm{i}}$ is
its value at the epoch when these electrons were accelerated. It is
natural to assume that the downstream magnetic field is frozen into the
gas. In this case it changes according to the relation $B_{\mathrm{d}} \propto a^{-2}$.
Let us assume that at the current epoch every downstream
volume element has $a \propto R_{\mathrm{s}}$. Then the expected decline of the 
synchrotron flux is 
\begin{equation}
  \frac{\mathrm{d}\lg S_{\nu}}{\mathrm{d}t} = -(2\gamma+3)
  \frac{V_{\mathrm{s}}}{R_{\mathrm{s}}},
\end{equation}
which gives $\mathrm{d}\lg S_{\nu}/\mathrm{d}t \approx -0.8$\%$/\mathrm{yr}$.
This value is in the observed range $\mathrm{d}\lg S_{\nu}/\mathrm{d}t \approx
-1.0$\%$/\mathrm{yr}$ to $-0.6$\%$/\mathrm{yr}$ (see Fig.\,\ref{Figure4}). We 
take also into
account that a revision of the data made be Rees (1990) led him to the
conclusion that the secular decrease of the flux of Cas\,A is about
$-0.8$\%$/\mathrm{yr}$ and is independent of frequency. This is consistent with
the recent result of \citet{Sullivan_MNRAS_1999_303} which gives
$-0.6$\%$/\mathrm{yr}$ at $\nu = 15\,\mathrm{GHz}$. We therefore conclude that
the observed relatively large secular decrease of the flux from Cas\,A is
consistent with the assumption \citep{Shklovsky_Supernovae_1968} that the flux
is dominated by the contribution of electrons, produced in a previous epoch,
which at the current stage undergo adiabatic cooling. Together with the
decrease of the downstream magnetic field this cooling causes the relatively
large rate of decrease of the radio flux.

The picture described is reproduced in our model (Fig.\,\ref{Figure4}). One can
see that the calculated secular decrease of the flux at $\nu \lsim
10\,\mathrm{GHz}$ is about $-0.5$\%$/\mathrm{yr}$ and thus very close to the
observations. At larger frequencies $\nu \gsim 10^3\mathrm{GHz}$ the
theoretically calculated decrease has a rather complicated dependence upon
$\nu$ and is on average larger than at low frequencies. The highest secular
decline predicted at the largest frequencies $\nu \gsim 10^{16}\mathrm{Hz}$ is
due to synchrotron losses which lead to the decrease of the electron maximum
energy.

Since the shock currently propagates through the free red supergiant wind, the
contribution of shell electrons to the total synchrotron flux will decrease in
time compared with that of the electrons produced by the shock in the free
wind. Therefore the rate of secular decline of the total synchrotron flux is
expected to decrease from the current value $-0.5$\%$/\mathrm{yr}$ towards
$-0.1$\%$/\mathrm{yr}$.

\subsection{The spatial structure of Cas\,A}

The dowstream spatial distribution of radio emitting electrons is
qualitatively consistent with the observed structure of Cas\,A. The
brightest part of the remnant in our model is the swept-up shell. It is
currently in the downstream region near the contact discontinuity $R_{\mathrm{p}}$
which separates ejecta and swept-up matter. We believe that it corresponds
to the observed bright radio ring. The `diffuse plateau' then corresponds
to emission coming from the swept-up free wind matter further out.

The small scale irregularities of the radio emission, the so-called compact
radio knots \citep[e.g.][]{Andersen_ApJ_1996_456} discussed earlier, which can
not be reproduced in our spherically symmetric model, could well be due to
amplification of the magnetic field at the bow shocks driven ahead of the
so-called fast moving knots, in the form of dense clumps of SNR ejecta.
Electrons with energies $\epsilon_{\mathrm{e}} < 1\,\mathrm{GeV}$, which
produce the radio emission, are strongly connected with the moving gas even on
such small scales because of their small diffusion coefficient. Therefore their
concentration also increases at the bow shock like the gas density.  Together
with the field amplification this leads to a strong increase of the radio
emission from these relatively small volumes.

We note that the field amplification leads to a decreasing effective energy
$\epsilon_{\mathrm{e}} \propto \sqrt{\nu/B}$ of the electrons which produce the
synchrotron emission at a given frequency $\nu$. Together with the adiabatic
electron heating, this leads to a brightness increase, and in parallel to a
steepening of the radio spectrum because of the concave shape of electron spectrum
in the corresponding energy range. Such a type of correlation between the knot
brightness and synchrotron spectral index is indeed observed
\citep{Andersen_ApJ_1996_456}.

The effect is expected to be larger in the outer downstream
region occupied by the swept-up free wind matter and recently accelerated
electrons compared with the swept-up shell region, where electrons after
their production have already adiabatically cooled by expansion of the
medium. Therefore a steeper synchrotron spectrum is expected from the
knots situated in the outer diffuse radio plateau surrounding the bright
ring, than from the knots in the ring. This is also indicated by the
observations
\citep{Andersen_ApJ_1996_456}.

\begin{figure}
  \centering
  \includegraphics[width=7.5cm]{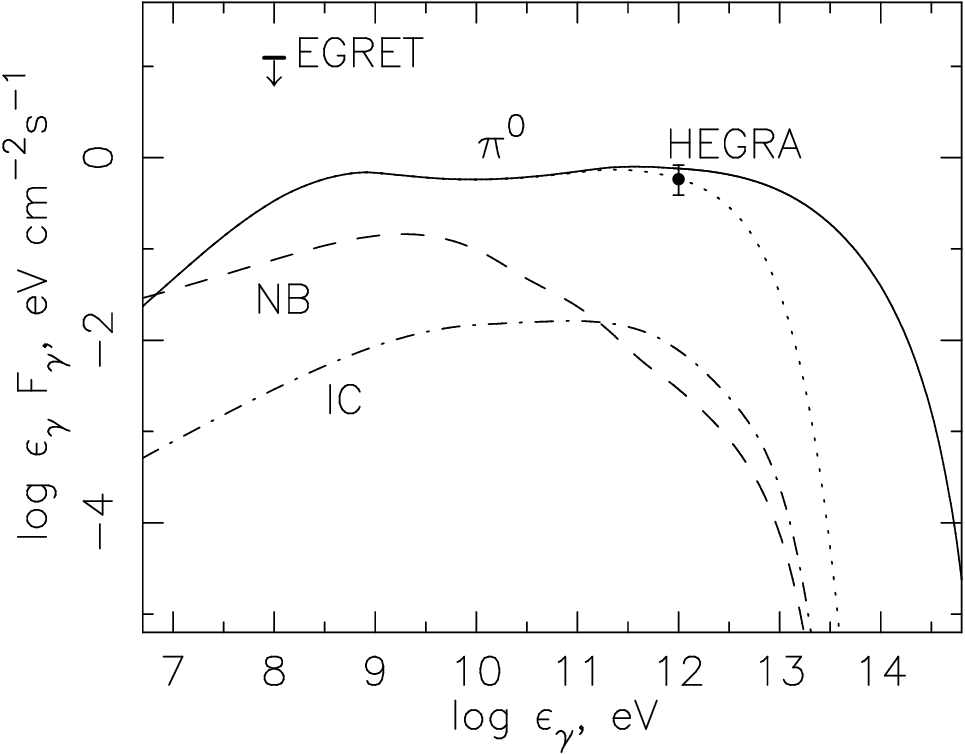}
  \caption{Inverse Compton (IC, {\it dash-dotted}), nonthermal Bremsstrahlung (NB, {\it dashed}), and
  $\pi^0$-decay ({\it solid}) integral $\gamma$-ray energy fluxes as a
  function of $\gamma$-ray energy. The dotted line represents the same
  $\pi^0$-decay \gr flux with an exponential cutoff $\exp
  (-\epsilon_{\gamma}/\epsilon_{\gamma}^{\mathrm{max}})$, where
  $\epsilon_{\gamma}^{\mathrm{max}} = 4\,\mathrm{TeV}$. 
  The $1\,\mathrm{TeV}$ data point is from HEGRA
  \citep{Aharonian_AA_2001_370}, the EGRET upper limit for 
  $\epsilon_{\gamma}>100$~MeV is from \citet{Esposito_ApJ_1996_461}.
  }
\label{Figure5}
\end{figure}

\subsection{High energy $\gamma$-ray emission from Cas\,A}

Figure~\ref{Figure5} represents the expected integral $\gamma$-ray 
energy flux components from
NB, IC scattering on the background radiation field 
(cosmic microwave + optical/infrared),
and hadronic collisions of CR protons with gas nuclei, respectively.

As already mentioned, in contrast with our spherically-symmetric model one
has to expect that not every part of the shock surface efficiently injects
and accelerates CRs. At the local portions of the shock, which are
currently quasiperpendicular the suprathermal particle injection is
presumably suppressed. Therefore the number of accelerated protons
calculated within the spherical approach should be corrected by some
renormalization factor $f_{\mathrm{re}} < 1$. It was argued earlier
\citep{Berezhko_AA_2002_395}
that values $f_{\mathrm{re}} = 0.15$ to $0.25$ are consistent with theoretical
considerations (V{\"o}lk et al., in preparation),
the average requirements of the
Galactic CR energy budget, and the observed structure of the remnant
SN\,1006. One of these arguments is that for a typical SNR our spherically
symmetric model predicts a significantly larger CR production $E_{\mathrm{c}} = 0.3$ to
$0.6 E_{\mathrm{sn}}$ than required for the Galactic energy budget
$E_{\mathrm{c}} \approx 0.1 E_{\mathrm{sn}}$ 
\citep[e.g.][]{Berezhko_Jetph_1996_82,Berezhko_Astropart_1997_7,Berezhko_AA_2000_357}.
The $\pi^0$-decay \gr
flux presented in Fig.\,\ref{Figure5} was calculated with a renormalization factor
$f_{\mathrm{re}} = 1/6$. Note that the number of accelerated electrons, which is required
for the observed synchrotron flux, remains the same. Therefore the actual
electron to proton ratio amounts to $K_{\mathrm{ep}} = 0.02$ instead of 
$4 \times 10^{-3}$. This also means that CRs absorb only $E_{\mathrm{c}} = 10^{49}\mathrm{erg}$
at the current epoch.

As one can see from Fig.\,\ref{Figure5}, 
at $\epsilon_{\gamma} = 1\,\mathrm{TeV}$ the $\pi^0$-decay energy flux
has a value of
$\epsilon_{\gamma} F_{\gamma} \sim 1\,\mathrm{eV}\,\mathrm{cm}^{-2}\mathrm{s}^{-1}$,
which is close to the experimental value of the reported integral TeV $\gamma$-ray flux
\citep{Aharonian_AA_2001_370}.
Also the slope of the $\pi^0$-spectrum lies in the reported range of allowed energy spectra,
which is quite large due to statistical uncertainties.
The $\pi^0$-decay energy flux hardens slightly above several GeV; nevertheless 
its value $\epsilon_{\gamma} F_{\gamma}$ remains practically constant between 300\,MeV and several TeV,
and has a cutoff energy of about 30\,TeV.

On the other hand,
the IC and NB fluxes at TeV energies 
are only at levels of
$\epsilon_{\gamma} F_{\gamma} \sim 10^{-2}\mathrm{eV}\,\mathrm{cm}^{-2}\mathrm{s}^{-1}$
and below.
Note that the electron scattering off the infrared/optical background
contributes about 50\% to the IC $\gamma$-ray flux at energies
$\epsilon_\mathrm{\gamma}<1\,\mathrm{TeV}$.
This is by a factor of two larger than the estimate of
\citet{Gaisser_ApJ_1998_492},
due to the fact that our electron spectrum is essentially steeper;
for such a spectrum, the contribution of background photons with higher energies
becomes more important.

The cutoff energy of the IC spectrum is about 2\,TeV.
At 30\,TeV, the leptonic $\gamma$-ray emission is entirely negligible.
The contribution from the CMB target field drops because of the cutoff of the electron spectrum;
the scattering off the IR/optical photons is suppressed as a result of the
decline of the Klein-Nishina cross section.
The cutoff energy of the IC spectrum resulting from the CMB photons is a bit lower than in the
case of IR/optical radiation. Therefore at $\epsilon_\mathrm{\gamma}>1\,\mathrm{TeV}$
the contribution of IR/optical background becomes progressively larger
compared to the CMB photons, so that at $\epsilon_\mathrm{\gamma}=10\,\mathrm{TeV}$ it exceeds
the CMB contribution by a factor of 200.

Figure \ref{Figure5} also shows the integral upper limit above 100\,MeV as reported by EGRET
\citep{Esposito_ApJ_1996_461}. This limit is well above the $\pi^0$-spectrum; even an energy-resolved
analysis of the EGRET data - which is not available in the literature yet - is not expected to yield
upper limits which would violate the model prediction.

Pion production by the accelerated nuclei leads not only to $\gamma$-rays but also to secondary
electrons and positrons. However, even for the large gas density
$N_{\mathrm{g}}\lsim50\,\mathrm{cm}^{-3}$ in the shocked shell, it can be easily shown that the 
ratio of secondary electrons to protons is more than two orders of magnitude lower than the
value $K_{\mathrm{ep}} = 0.02$ for primaries deduced here. Therefore the contribution of secondaries to
the synchrotron and Inverse Compton emission and the nonthermal Bremsstrahlung is negligible.

\subsection{Remaining model uncertainties}

We believe that there is no significant difference between the calculated and
measured TeV \gr flux of Cas\,A.  Nevertheless we have to ask ourselves to
which extent our calculated \gr flux is a robust result and which effects could
modify it.

This concerns first of all the magnetic field configuration in the shell and in
the free wind region. The RSG wind is most probably driven by strong wave
activity from the star's outer convection layers, essentially Alfv\'en waves
\citep{Hartmann_ApJ_1980_242}.


Their wavelengths should be quite large and unrelated to the gyro radius scale
of the accelerated particles. The SNR shock propagates through this turbulent
wind and injection occurs at the quasi-parallel portions of the magnetic field
lines. For large enough wave amplitudes in the wind, the mean field direction
is no more relevant for the acceleration in a strong shock that produces in
addition its own large-amplitude resonantly scattering waves
\citep{Lucek_MNRAS_2000_314}. In the shell the situation is even more
pronounced because of the compression by the Wolf-Rayet wind from behind and the
resulting hydrodynamic instabilities \citep{Garcia-Segura_AA_1996_316}. We
therefore believe that it is justified to use the Bohm limit for the diffusion
coefficient, renormalizing the flux since the injection at the
instantaneously quasi-perpendicular shock positions is reduced. Thereby we neglect the
possible systematic effect of the mean field that in principle has a Parker
spiral type configuration with the tendency to further reduce the injection
rate.

The second physical effect which can play a role in the current
evolutionary phase is CR escape
during the more recent phase of the SNR evolution. 
Let us assume that the red supergiant wind speed 
$V_{\mathrm{w}} \approx 10\,\mathrm{km/s}$ does not exceed the 
Alfv\'en speed, and that the field strength in the free red supergiant wind 
$B_{\mathrm{w2}}$ is as low as $10\,\mu\mathrm{G}$.
In this case the maximum momentum of accelerated protons
\begin{equation}
  p_{\mathrm{max}} \propto R_{\mathrm{s}}V_{\mathrm{s}} / \kappa(mc)
\end{equation}
is at least 20 times smaller compared with its value in the shell since
the minimum CR diffusion coefficient $\kappa \propto 1/B$ increases by a
factor of 20. In such a situation one should expect that the diffusion
coefficient of particles with $p > p_{\mathrm{max}}^{\mathrm{w}}$,
where $p_{\mathrm{max}}^{\mathrm{w}}$ is the
maximum momentum of CRs accelerated in the red supergiant wind region
$r > R_2$, exceeds the Bohm limit because the shock becomes unable to
accelerate them and they consequently do not produce high level Alfv\'enic
turbulence near the shock. It is like if the shock had become "old",
being unable to confine the highest energy particles it accelerated
previously in the shell. Therefore these particles leave the SNR very soon
after the SN shock reaches the red supergiant wind region, resulting in a
much smaller production of $\pi^0$-decay $\gamma$-rays with the highest
energies compared with the previous cases. For the electron X-ray emission
this is a small effect since their main synchrotron radiation comes from
the shell; their IC emission remains essentially the same with or without
escape.

To illustrate this scenario we present in Fig.\,\ref{Figure5} a \gr spectrum which compared
to the previous one has an additional cutoff factor
$\exp(-\epsilon_{\gamma}/\epsilon_{\gamma}^{\mathrm{max}})$; the cutoff energy
$\epsilon_{\gamma}^{\mathrm{max}}$ was set to $4\,\mathrm{TeV}$ which corresponds to the escape of
particles with energies above $50\,\mathrm{TeV}$, lowering the proton maximum energy
by a factor of 20. As one can see from Fig.\,\ref{Figure5} this would still be
consistent with the HEGRA flux, but steepen the spectrum at $1\,\mathrm{TeV}$.

The integral NB and IC energy fluxes are almost 2 orders of magnitude
lower than the corresponding $\pi^0$-decay flux. Even taking into account
the local thermal far infrared fluxes can not increase the IC flux by more
than a factor of 2 
(R.J. Tuffs, private communication). 
Therefore the dominance of the $\pi^0$-decay flux is a rather robust result.

\section{Summary}

Perhaps the most important result of our considerations is that the spectral
shape of shock accelerated electrons with their essential synchrotron cooling
in the downstream region is very well consistent with the observed synchrotron
emission. To reproduce a very steep radio spectrum $S_{\nu} \propto
\nu^{-0.77}$ the shock must be strongly modified. This shock modification can
only be produced by accelerated protons if they are also efficiently injected
into the acceleration process, as it was assumed in the calculation.

The significant synchrotron losses of electrons in the strong interior
magnetic field makes their spectrum steep $N_{\mathrm{e}} \propto \epsilon_{\mathrm{e}}^{-3}$
also at high energies $\epsilon_{\mathrm{e}} > 10\,\mathrm{GeV}$. This leads to a flat connection
of the spectral energy distributions of the observed radio and X-ray
synchrotron emissions.

The rather high secular decline of the synchrotron radiation observed in
the radio range is naturally reproduced in our model,
because this range is dominated by electrons which were accelerated at a previous epoch 
and undergo currently adiabatic and synchrotron cooling in the expanding downstream region.

We find that after reduction of the predictions of the nonlinear
spherically-symmetric model by a renormalization of the number of
accelerated nuclear CRs, to take account of the large areas of
quasiperpendicular shock regions of a SNR, good consistency with all
observational data can be achieved, including the reported TeV
$\gamma$-ray flux 
\citep{Aharonian_AA_2001_370}.
The used renormalisation factor is consistent with the need that
the average acceleration efficiency of a typical SNR
within our model scenario should meet the requirements for Galactic CR
acceleration.

In addition, our calculations show that at all energies above $1\,\mathrm{GeV}$
the \gr production is dominated by $\pi^0$-decay. At TeV energies the expected
$\pi^0$-decay flux exceeds the IC and NB fluxes by a factor of about seventy.
Therefore the leptonic emission is totally inadequate to explain the observed
TeV $\gamma$-ray flux. The $\pi^0$-decay spectrum $F_{\gamma}^{\pi} \propto
\epsilon_{\gamma}^{-1}$ extends up to $30\,\mathrm{TeV}$, whereas the IC and NB
$\gamma$-ray fluxes have a cutoff at about $1\,\mathrm{TeV}$. Therefore the
detection of $\gamma$-ray emission at $10\,\mathrm{TeV}$ and above would imply
further evidence for its hadronic origin.

We conclude that the observed properties of the radio and X-ray emission
can be explained within the assumption that the SN blast wave is the main
source of energetic particles in Cas\,A. The CR production efficiency
and the electron to proton ratio implied by these multi-wavelength
observations are consistent with the requirements of the nuclear CR sources in
the Galaxy.

\begin{acknowledgements}
This work has been supported in part by the Russian Foundation for Basic
Research (grants 00-02-17728, 99-02-16325) and by the Russian Federal
Program "Astronomiya" (grant 1.2.3.6). EGB acknowledges the hospitality
of the Max-Planck-Institut f\"ur Kernphysik where part of this work was
carried out. The authors thank F. Aharonian, W. Hofmann and R. Tuffs for
valuable discussions.
\end{acknowledgements}

\bibliographystyle{aa.bst}
\bibliography{H4000}

\end{document}